\documentclass[prl,twocolumn,superscriptaddress]{revtex4}
\usepackage{epsfig}
\usepackage{times}
\usepackage{amsmath}
\usepackage{amsfonts}
\usepackage{amssymb}
\usepackage[usenames]{color}
\usepackage{graphicx}

\newcommand{\beq}{\begin{equation}}
\newcommand{\eeq}{\end{equation}}
\newcommand{\beqa}{\begin{eqnarray}}
\newcommand{\eeqa}{\end{eqnarray}}

\begin{document}
\title{Distributed quantum information processing with mobile electrons}
\author{Yuichiro Matsuzaki}
\affiliation{Department of Materials, University of Oxford, OX1 3PH,
U. K.}
\author{John H. Jefferson}
\affiliation{ Department of Physics, Lancaster University, Lancaster LA1 4YB, UK
}

\begin{abstract}
 In distributed quantum information processing,
 small devices composed of a single or a few qubits
  are networked together through shared entanglement to achieve a scalable machine.
 Typically,
 photons are utilized to generate remote entanglement between optically active matter qubits.
In this paper, we consider another possibility for achieving
 entanglement between nodes: using 
mobile electron spins as mediators of the interaction between
static qubits at each
node.
  A strong
 interaction between electrons makes it feasible to couple the flying
 electron with the static electron.
However, since the electrons easily interact with the environment, 
 error accumulation
during the entanglement operation could be more severe
 than with the other strategy using photons as flying qubits.
We introduce a new scheme especially designed to minimize such error accumulation by
using several distillation protocols.
The conclusion is that a high fidelity entanglement operation can be constructed even
under the effect of typical imperfections,
and this suggestion therefore offers a feasible route for the realization of
distributed quantum information processing in solid state systems.
\end{abstract}

\maketitle

The problem of scalability is a key challenge for practical
  quantum information.
  Recently, to overcome this problem, a new approach called
 ``distributed quantum information processing'' was suggested
 \cite{CEHM01a,BK01a,BPH01a,BKPV01a,FZLGX01a,UTprl}.
 In this scheme, spatially separated  qubits are entangled 
 to make a network between
  small devices composed of a single or a few qubits.
 The distance
 between qubits in this scheme makes it easy to
 address the individual qubits and suppress decoherence caused by unknown
 interaction between them. So this scheme is considered to 
 scale well for a large number of the qubits.
To construct a high-fidelity entanglement operation (EO),
many proposals using a photon
to mediate interactions
between the matter qubits have been suggested
  \cite{CEHM01a,BK01a,BPH01a,BKPV01a,PRAmatsuzaki2010distributed},
  because a photon is almost decoherence free due
to the weak interaction with the environment and therefore is considered
as an ideal flying qubit.

On the other hand, there is another possibility for a flying qubit: a
mobile electron spin in solid state systems.
The ability to coherently control an electron is at the heart of
recent developments.
 A strong
interaction between electrons makes it feasible to interact flying
electron spins
with static electron spins, so that
distant static electron spins
can exchange information with the help of the flying electron spins.
So
 the investigation of flying electron qubits is an area of active study.
A typical experimental setup
is that
a flying qubit injected in a one dimensional system passes
between two distant matter qubits in order to
mediate entanglement between them,
and  EO protocols
in this setup
have been investigated by many
authors
\cite{costa2006entanglement,ciccarello2006entanglement,habgood2009scattering,ciccarello2008extraction,yuasa2009efficient,habgood2008entanglement}.
However,
 in such solid-state systems, the effect of error
accumulation during the EO becomes more
relevant than in the case of an optical flying qubit.
If one uses an electron as a flying qubit, decoherence
from the environment is inevitable, which results in
the degradation of the entanglement. 
In addition, controlling
interaction between the flying qubit and static qubit is still
challenging with current technology, and imperfection of the control
will be another source of errors.
   Although  certain
 levels of error can be subsequently dealt with through techniques such
 as 
 quantum error corrections
 \cite{raussendorf2007fault},
     keeping the errors as low as possible is still essential in order to make such techniques
   feasible.


In this paper, we suggest a new form of EO between
matter qubits through a flying electron qubit. This scheme is especially
designed to minimize the error accumulation through
 the method called distillation.
By using
 imperfect entangled states as resources, one can generate a high
 fidelity entanglement through distillation protocols.
In the (unrealistic) assumption of no decoherence and fine tuned parameters, our
protocol provides a way to make a perfect Bell pair between static qubits
with unit
probability without measuring the flying qubit, while most of
the previous schemes are probabilistic or require projective measurements
to the flying
qubit
\cite{costa2006entanglement,ciccarello2006entanglement,habgood2009scattering,ciccarello2008extraction,yuasa2009efficient,habgood2008entanglement}.
Since such measurements on the flying qubit are difficult to
realize, our scheme may prove more feasible than the previous ones.
Also, we estimate the error
accumulation during the EO
due to decoherence from the environment and typical experimental
imperfections, 
and we show how to recover the fidelity by
using distillation protocols.
Since distillation protocols require more than two qubits per
node, we assume that a few ancillary qubits are located near the static qubit.
We show that even adding only one ancillary qubit at each node
makes the EO robust against typical experimental imperfections.

\begin{figure}[h]
     \begin{center}
         \includegraphics[width=7cm]{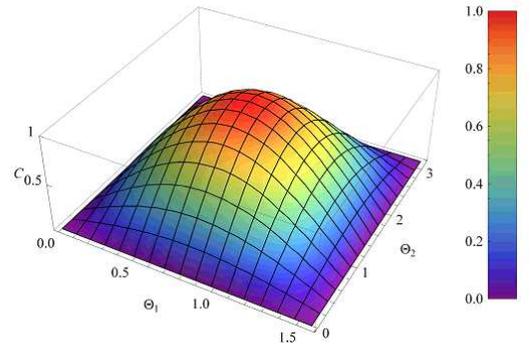}
     \end{center}
                \caption{A concurrence of the entanglement
                between the
                static qubits 
                is plotted.
                Here, $\theta _1$ ($\theta _2$) denotes the phase shift
                induced by the interaction between the
                first (second) static qubit and the flying qubit. }\label{con-static}
     \end{figure}

 We consider a
 two-electron scattering model in a one dimensional
 system where one electron propagates 
  and the
other electron is in the ground state of a confining potential,
 introduced in \cite{jefferson2006entanglement}.
  For example, an electron injected into the conduction band of a carbon nanotube
and an electron bounded by a potential well in the same band 
can be used as
the flying and the static
qubits \cite{gunlycke2006entanglement} respectively.
One can construct a quantum well between
electrodes in the nanotube, and the energy levels of the well are discrete
due to the high degree of confinement in all three dimensions, which forms a quantum dot to accommodate the static qubit.
Note that the relevant interaction between the electrons arises from
 Coulomb repulsion which has no dependency on the spins of the
 electrons and may be described by the effective Hamiltonian \cite{jefferson2006entanglement}:
\begin{eqnarray}
 H=\frac{1}{2m}\hat{p}_a^2+ \frac{1}{2m}\hat{p}_b^2+\hat{v}(x_a)+\hat{v}(x_b)
  +\hat{V}(x_a-x_b), \label{john-hamiltonian}
\end{eqnarray}
where $\hat{v}(x)$ is an effective one electron
potential, $\hat{V}(x_a-x_b)$ is an effective two
electron potential, and $m$ denotes the effective mass of the electron.
 Since the total wave function
 of fermions should 
 be anti-symmetric, 
the spin state becomes symmetric for an anti-symmetric
spatial wave function while the spin state becomes anti-symmetric for a symmetric
spatial wave function.
So, even for the spin-independent Hamiltonian, the scattering process depends on the spin states.
of the two-body quantum systems.
For the spin states of two electrons, there are three symmetric states (triplet)
 and one anti-symmetric state (singlet).
The singlet is represented as
$  |S\rangle =\frac{1}{\sqrt{2}}|\uparrow \downarrow\rangle -
  \frac{1}{\sqrt{2}}|\downarrow \uparrow\rangle $
and the triplets are represented as $|T_1\rangle =|\uparrow
\uparrow \rangle $, $|T_0\rangle =\frac{1}{\sqrt{2}}|\uparrow \downarrow\rangle +
  \frac{1}{\sqrt{2}}|\downarrow \uparrow\rangle$, and $|T_{-1}\rangle =|\downarrow \downarrow \rangle $.
If the kinetic
 energy of the initial flying qubit is smaller than the energy difference
 between the ground state and the first excited state of the static
 qubit, the quantum state of the flying qubit after the scattering has the same
 magnitude of the momentum as the initial state, while the
 static qubit remains in the ground state.
 The state of the flying qubit after the scattering becomes a superposition of a reflected state and a transmitted state 
and, due to the Pauli exclusion principle, the amplitudes of
reflection and transmittance depend on the spin states.
Furthermore, in this model and its extension to multiple static and
 propagating electrons, the total number of up spins
and down spins are
 conserved
 throughout the interaction, reflecting conservation of
 total magnetization imposed by the Hamiltonian
 (\ref{john-hamiltonian}). 
 Therefore, for the initial state $|\uparrow \downarrow \rangle _i=\frac{1}{\sqrt{2}} |T_0\rangle _i
   +\frac{1}{\sqrt{2}}|S\rangle _i$,
 we obtain \cite{jefferson2006entanglement} 
 \begin{eqnarray}
 U |\uparrow \downarrow \rangle _i
   &= &\frac{1}{\sqrt{2}}(
    r_T |T_0\rangle _{r}+ t_T|T_0\rangle _{t}+r_S|S\rangle _r+t_S|S\rangle _t
    )\nonumber 
 \end{eqnarray}
 where $U$ denotes a unitary evolution by the Hamiltonian, $|\rangle
_r$ ($|\rangle _t$) denotes the state of the reflected (transmitted)
flying qubit, and $|\rangle _i$ denotes the initial state before the
scattering. 
Also, $r_S$ and $t_S$ ($r_T$ and $t_T$) are complex numbers
to denote the
amplitudes of the reflection and transmittance when the initial state is
a singlet (triplet).
  These amplitudes can be determined by solving the
  spin-independent scattering problem where the orbital wave
  functions must be either symmetric
                                or asymmetric.
                                 This two-electron system may be constructed such that
 a singlet is on resonance (with $|t_S|=1$)
 while the triplet is off resonance (with $|t_T|\simeq 0$).
This may be achieved by choosing the shape of the
 potential to give a large separation between singlet and triplet
 resonances and adjusting biases to bring the singlet on resonance
 \cite{jefferson2006entanglement, gunlycke2006entanglement}.
 In this case,
 if one were to projectively measure whether the flying qubit had been
 transmitted or not via charge detection, then
 the spin state between the transmitted flying
  qubit and the static qubit would be a near perfect Bell pair.
The success probability to observe such transmission is $50\%$
 because the reflection
 probability becomes precisely equal to the transmission probability
 at this resonance
in the limit of perfect triplet
 blocking, $t_T=0$.
Also, we study a case of forward scattering where $r_S,r_T\simeq 0$.
 We have  $U|S\rangle _i\simeq t_S|S\rangle _t= e^{i\theta
 _S}|S\rangle _t$ and $U|T\rangle _i\simeq t_T|T\rangle _t= e^{i\theta
 _T}|T\rangle _t$. So we obtain $ U|\uparrow \downarrow \rangle
 _i\simeq e^{i\theta '}(\cos \theta |\uparrow \downarrow \rangle _{t}+i\sin
   \theta |\downarrow \uparrow \rangle _t)$
 where $\theta '=\frac{\theta
 _T +\theta _S}{2}$ and $\theta =\frac{\theta _T-\theta _S}{2}$.
 Similarly, we have $ U|\downarrow \uparrow \rangle _i\simeq e^{i\theta '}(i\sin \theta |\uparrow \downarrow \rangle _{t}+\cos
   \theta |\downarrow \uparrow \rangle _t)$ and so, when the spins are anti-parallel, the quantum state after the interaction
 becomes a superposition of the non-spin flip state and spin-flip state.
 Furthermore, we have $  U|\uparrow \uparrow \rangle _i\simeq
   e^{i(\theta +\theta ') }|\uparrow \uparrow
   \rangle _t$ and $U|\downarrow \downarrow \rangle _i\simeq
     e^{i(\theta +\theta ')}|\downarrow \downarrow \rangle _t$.
 and so spin flip processes do not occur when the
 initial spins are parallel.
      We now show how two static qubits are entangled through an
   interaction mediated by a
   flying qubit.
   We begin by describing a simplified case with fine
   tuned parameters and no decoherence and consider realistic imperfections later.
   Suppose that two static qubits (bound electrons) $s1$ and $s2$ are located in a one dimensional
   system. We assume that, due to the distance between them, the interaction between the
   static qubits is negligible. 
   We prepare an initial state $|\uparrow \downarrow \downarrow \rangle
   _{f,s1,s2}$ and we will send a flying qubit as a mediator between the
   static qubits. 
   For the forward scattering, we obtain
   \begin{eqnarray}
    U_2U_1 |\uparrow \downarrow \downarrow \rangle _{f,s1,s2}
    =e^{i(\theta '_1+\theta '_2)}(\cos \theta _1\cos \theta _2|\uparrow
     \downarrow \downarrow \rangle _{f,s1,s2}
     \nonumber \\
     +i\cos \theta _1 \sin \theta _2|\downarrow \downarrow \uparrow
     \rangle _{f,s1,s2}
     +ie^{i\theta _2}\sin
     \theta _1|\downarrow \uparrow \downarrow \rangle _{f,s1,s2})
     \label{total-state}
   \end{eqnarray}
   where $U_j$ $(j=1,2)$ denotes a unitary operation for the interaction
   between the $j$th
   static qubit and the flying qubit.
   Note that, even when there is a finite reflection probability during
   the interaction between the flying and the first static qubit, one
   can get the same form via a postselection of transmission by using
   a charge detection.
   Since we are only interested in the quantum states of the static qubits,
   we trace out the flying qubit and obtain
   \begin{eqnarray}
 \rho _{s1,s2}=(1-\frac{P_1+P_2}{2})|\downarrow \downarrow \rangle
  _{s1,s2}\langle \downarrow \downarrow |
    + \frac{P_1+P_2}{2} |\chi \rangle \langle \chi | \ \ \ 
  \label{earl-state-p}\ \ \ 
\end{eqnarray}
where $  |\chi \rangle =\sqrt{\frac{P_1}{P_1+P_2}}|\downarrow \uparrow \rangle
  _{s1,s2}+ \sqrt{\frac{P_2}{P_1+P_2}}ie^{i\theta _2 }|\uparrow \downarrow \rangle
  _{s1,s2}$, $P_1=2 \cos ^2 \theta _1 \sin ^2 \theta
   _2$, and $P_2= 2\sin ^2 \theta _1$.
We can calculate the the concurrence of this state as $C=\sqrt{P_1P_2}$.
We have plotted this concurrence in the FIG.\ref{con-static}.
and obtain a perfect Bell state for $\theta _1=\frac{\pi
}{4}$, $\theta _2=\frac{\pi }{2}$, as can be seen
directly from (\ref{total-state}). For this choice,
a Bell pair between the first static qubit and the flying qubit is generated
while
the latter interaction plays a role of a SWAP gate between the flying
qubit and the second static qubit. Surprisingly, 
a perfect Bell pair can be deterministically generated without any
measurements, while most of the previous schemes are probabilistic or
require spin-resolving measurements on
the flying qubit
\cite{costa2006entanglement,ciccarello2006entanglement,habgood2009scattering,ciccarello2008extraction,yuasa2009efficient,habgood2008entanglement}.
One of the ways to construct such gates is as follows.
As already mentioned, for suitable parameters of the
potential,
one can make the energy of the singlet resonance far from
that of the triplet resonance
\cite{jefferson2006entanglement,gunlycke2006entanglement}
so that a Bell pair
between the flying and the static qubits can be generated for a
particular kinetic energy
in a weak to intermediate correlation regime.
 This regime is suitable for our first gate between the flying and
 the static qubits.
 Conversely, 
in a strong correlation regime (wide dot) where electrons avoid each other to
lower their Coulomb repulsion energy
in
the quantum well,
transmission can be almost unity for both the singlet and the triplet 
 \cite{gunlycke2006entanglement}.
 This is 
 due to the small single-triplet energy splitting and
 a small overlap between the wave functions of the
 flying and the static qubits in the quantum dot.
Moreover, since the bound electron feels the Coulomb repulsion of the incident
electron as it enters the dot,
the bound electron is
ejected
and the incident electron becomes bound in the large quantum dot.
Since this process
represents an
exchange between the static and flying qubits,
an effective SWAP gate with almost unity transmission probability
 is performed in this process  \cite{gunlycke2006entanglement}, which is suitable for our second gate.
Although a reflection probability at the first gate
is
high ($\sim 50\%$),
such reflection error can be heralded
by using a charge detector such as a single electron transistor
\cite{schoelkopf1998radio}
to check whether
the flying qubit is transmitted.

\begin{figure}[h]
    \begin{center}
        \includegraphics[width=6cm]{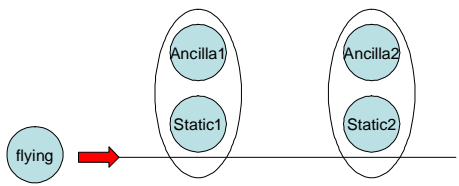}
    \end{center}
   \caption{Schematic of a construction of a high fidelity parity
   projection between distant static qubits where each node
   supports two qubit storage. Transmission of the flying qubit could
   generate
   only impure entanglement between the static qubits due to
   experimental imperfections.
   One can use the additional ancillary qubits
   for a distillation protocol to guarantee that an ideal EO is performed.
   }\label{earl-static-fly}
    \end{figure}

In the actual implementation of the above scheme,
a perfect Bell pair might not be generated due to imperfect
control of the interaction between the static and flying qubits.
 In order to overcome such difficulty of controlling parameters, we adopt a two-step distillation protocol, which requires
 an ancillary qubit per node (See FIG. \ref{earl-static-fly}).
 Since we
 expect that the primary error sources are associated with the flying
 qubit,
 we assume that high
 fidelity local operations
 are possible between the static qubits
 within each node. We will show that adding only one ancillary
 static qubit makes our scheme robust against imperfections.
   After the flying qubit has passed
   the two static qubits, 
   we obtain the state $\rho _{s1,s2}$ 
    which is of the same form as the key state considered in \cite{CB01a,PRAmatsuzaki2010distributed}.
It follows that, if we have an ancillary qubit near the static qubit at each location, we can perform an
efficient two round distillation protocol.
 We utilize 
   the state $\rho _{s1,s2}$  as a resource to perform a parity projection
   on the ancillary qubits.
Since the state $\rho _{s1,s2}$ is mixed, the parity projection is also
   imperfect.
   However, we can generate the state $\rho _{s1,s2}$ again through a
   transmission of another flying qubit and can utilize this state for
   performing the second parity projection on the ancillary qubits. From the measurement results
   of the two parity projections, we know whether
   the ancillary qubits are projected onto an entangled state
   or 
   just projected onto separable states.
As long as the errors caused by local operations in the node are
negligible, this protocol constructs a perfect parity projection
with
success probability $P_s=\frac{P_1P_2}{2}$.
Although errors in the local operations reduce the fidelity of the
parity projection, it has been shown that this does not represent a major
issue because only a few operations are necessary in this protocol \cite{CB01a}.
Note that parity projection is one of the most commonly proposed
  EOs \cite{BK01a, CB01a,PRAmatsuzaki2010distributed,UTprl}
  to make useful multipartite entanglement
  such as
  a cluster state
   for quantum
   computation \cite{Raussendorf:2001p368} and a GHZ state for a
   quantum magnetic sensor \cite{huelga1997improvement}.

By using a projective measurement and a subsequent single qubit rotation,
one can
prepare any initial spin states.
However, for a flying qubit, it would be difficult to construct a
high fidelity measurement with the current technology
and thus the initialization of the flying qubit
could be imperfect.
Importantly, this imperfection can be detected through the two round
protocol and prevented from reducing the fidelity.
Due to the imperfect initialization of the flying qubit, the initial
state of the flying qubit can be represented as $(1-\epsilon )|\uparrow
\rangle _f\langle \uparrow |+\epsilon |\downarrow \rangle _f\langle
\downarrow |$.
Through the transmission of the flying qubit between the static qubits,
the state of the static qubits will be $ \rho ^{(\epsilon )}_{s1,s2} =
(1-P')|\downarrow
  \downarrow \rangle _{s1,s2}\langle \downarrow \downarrow |+ P'|\chi
  \rangle \langle \chi |$ 
where $|\chi \rangle =\sqrt{\frac{P_1}{P_1+P_2}}|\downarrow \uparrow
\rangle _{s1,s2}+ie^{i\theta _2 }\sqrt{\frac{P_2}{P_1+P_2}}|\uparrow
\downarrow \rangle $ and $P'=(1-\epsilon)\frac{P_1+P_2}{2}$.
Therefore, one can perform a high fidelity parity projection between the
ancillary qubits through the two round protocol.
The success probability of this distillation is $P^{(\epsilon )}_s=\frac{1}{2}(1-\epsilon )^2P_1P_2$,
which means that the imperfection of the initialization just decreases the success
probability of the EO without affecting the fidelity
of the target entanglement.

Dephasing and relaxation are relevant error sources in solid state systems.
 Especially when the distance between the static qubits is long, the
 flying qubit will be affected by such errors
 due to the noisy
 environment.
 However, similar to the imperfect initialization, the relaxation
 process, which makes a spin-up state
 into a spin-down state,
 only decreases the success probability without affecting the fidelity
of the EO.
 Therefore, we study especially how dephasing the flying qubit affects the
 fidelity of the EO, and also
 introduce a way to reduce the impact of this problem.
We adopt a
 general dephasing model such that the state is acted on by a Pauli
 matrix $\hat{\sigma }_z$
 with a probability $\epsilon
 _{\text{z}}$ otherwise the
 state is unchanged.
 Note that
 dephasing does not
change the initial state before the flying qubit interacts with the static
 qubits, and
 also
 the flying qubit will be traced out after the flying qubit has passed the two static qubits.
 So the dephasing effect is
 only important when the flying qubit is between the static qubits.
The final state of the static qubits is then
 represented as $ \rho '_{s1,s2}=(1-\epsilon _{\text{z}})\rho _{s1,s2}+\epsilon
  _{\text{z}}\hat{\sigma }_{z}^{(s1)}\ \rho _{s1,s2}\ \hat{\sigma }_{z}^{(s1)}$
from (\ref{earl-state-p}).
 For an initial state $|++\rangle  _{\text{a1,a2}}$ of the ancillary
 qubits, we can perform the two round protocol by using the state
 $\rho '_{s1,s2}$, and we obtain 
 $   \rho _{\text{a1,a2}}\simeq (1-2\epsilon _{\text{z}})|\psi
    ^{(+)}\rangle \langle \psi ^{(+)}|
    +2\epsilon _{\text{z}}|\psi
    ^{(-)}\rangle  \langle \psi ^{(-)}| $
  where $|\psi
   ^{(\pm )}
   \rangle =\frac{1}{\sqrt{2}}|\uparrow \downarrow \rangle \pm
   \frac{1}{\sqrt{2}}|\downarrow \uparrow \rangle $.
 This means that, due to the
 dephasing, the fidelity of the EO decreases to
 $1-2\epsilon _{\text{z}}$.
 Fortunately, an
 efficient distillation protocol to reduce the effect of the
 dephasing has been suggested \cite{C01a}, and this protocol requires
 only an additional qubit at each node.
 So, if we have a second ancillary qubit, 
 we
 can 
 reduce the impact of the dephasing as follows.
   First, we prepare the imperfect bell state $\rho _{\text{a1},\text{a2}}$ between
   the first ancillary qubits by the two round protocol.
   Since this two round protocol is probabilistic, we repeat this process
   until successful.
   Second, we perform local operations (including measurements) between
   the first and second ancillary qubits at each node.
   By repeating the first and second steps, the state of the second
   ancillary qubits converges to a high fidelity entangled state.
   In this process, one obtains a series of
   measurement results. The fidelity of the EO becomes a function of the
   measurement results and  follows a biased random walk which 
   converges rapidly to unity \cite{C01a}.
   For example, with  a dephasing rate $\epsilon
   _{(\text{z})}=0.089$, one can obtain a high fidelity $F\simeq
   1-10^{-4}$ entangled state
 by performing this protocol $10$ times on average \cite{C01a}.
 Importantly, this protocol does not require postselection and hence one
 can perform an EO deterministically
 with no risk of damaging previous entanglement for performing a new
 EO.
 So this EO has the advantage
   of reducing the time resource needed \cite{UTprl}
 especially when
 one tries to generate multipartite entanglement.
 
 In order to generate a large entangled state, one has to prepare
 a significant number of
 static qubits in the one dimensional system and needs to perform
 EOs between two specific static qubits \cite{BK01a}.
 Here, we describe how to perform such selective EOs via a flying qubit.
 Suppose that one prepares a state $|\uparrow \rangle _{f} \Big{(}\bigotimes
 _{j=1}^{i-1}|\uparrow \rangle _{s_j}\Big{)} |\downarrow \downarrow
 \rangle _{s_i,s_{i+1}}
 \Big{(}\bigotimes
 _{m=i+2}^{N}|\uparrow \rangle _{s_m}\Big{)} $ and try to perform an EO
 between the down-spin static qubits at $i$ and $i+1$.
 For the triplet resonance,
 the flying qubit $|\uparrow \rangle _f$
 has spin-preserving, unity transmission probability with a static
 qubit $|\uparrow \rangle _{s}$
 \cite{jefferson2006entanglement}.
 Hence, the flying qubit arrives at the $i$ th static
 qubit without affecting the previous static qubits.
 After the flying
 qubit interacts with the $i$ th and $i+1$ th static qubits,
 the flying qubit can continue to transmit until
 it reaches the end of the one dimensional system through the
 SWAP gate 
 with a unity transmission probability in the strong correlation regime.
 As a result, one obtains the target state
 $\rho _{s1,s2}$ between the $i$ th and $i+1$
 th static qubits and, by using this state, one can perform an EO to the ancillary qubits.

Finally, we discuss a possible experimental realization for the
distillation protocols.
Molecular spin
systems could be attached externally to the nanotube
to provide
ancillary qubits.
For example,  Sc@C${}_{82}$ has an unpaired electron spin in the highest
occupied molecular orbital
with a long
 decoherence time of 
 $200$ $\mu s$, which can be
 utilized as a qubit
                     \cite{richard2010coherenceball}.
  Moreover, since the electron spin exists mainly on the the fullerene cage,
  there is expected to be a large exchange
  coupling between
 the electron spin and the static qubit in the quantum dot \cite{ge2008modeling}.
 Furthermore, instead of using a quantum dot 
 for a static qubit,
  Sc@C${}_{82}$ could itself
 provide both the static qubit and the ancillary
 qubit, the latter being provided by a nuclear spin.
This would allow two-qubit gate operations between the nuclear and the
  electron spins through hyperfine coupling
  \cite{benjamin06}.
  So these properties of Sc@C${}_{82}$
  may be
 suitable for performing our distillation protocols, although
 further research is needed to assess its suitability as a future
  generation technology.
In conclusion, we have suggested a way to construct a high fidelity
EO between spatially separated static qubits by using a flying qubit in
solid state systems.
Although there are several types of error, 
we have shown a robust way to perform a high-fidelity EO
by making use of a distillation protocol, suggesting constitutes a feasible route for the realization of
distributed quantum information processing in solid state systems.
Authors thank Simon C. Benjamin and Joseph Fitzsimons for useful discussions.




\end{document}